# Portfolio Optimization on NIFTY Thematic Sector Stocks Using an LSTM Model

Jaydip Sen
Department of Data Science
Praxis Business School
Kolkata, INDIA
email: jaydip.sen@acm.org

Saikat Mondal
Department of Data Science
Praxis Business School
Kolkata, INDIA
email: saikatmondal15@gmail.com

Sidra Mehtab
Department of Data Science
Praxis Business School
Kolkata, INDIA
email: smehtab@acm.org

*Abstract*— Portfolio optimization has been a broad and intense area of interest for quantitative and statistical finance researchers and financial analysts. It is a challenging task to design a portfolio of stocks to arrive at the optimized values of the return and risk. This paper presents an algorithmic approach for designing optimum risk and eigen portfolios for five thematic sectors of the NSE of India. The prices of the stocks are extracted from the web from Jan 1, 2016, to Dec 31, 2020. Optimum risk and eigen portfolios for each sector are designed based on ten critical stocks from the sector. An LSTM model is designed for predicting future stock prices. Seven months after the portfolios were formed, on Aug 3, 2021, the actual returns of the portfolios are compared with the LSTM-predicted returns. The predicted and the actual returns indicate a very high-level accuracy of the LSTM model.

## I. Introduction

The design of optimized portfolios has remained a research topic of broad and intense interest among the researchers of quantitative and statistical finance for a long time. An optimum portfolio allocates the weights to a set of capital assets in a way that optimizes the return and risk of those assets. Markowitz in his seminal work proposed the mean-variance optimization approach which is based on the mean and covariance matrix of asset returns [1]. Despite the elegance in its theoretical framework, the mean-variance theory of portfolio has some major limitations. One of the major problems being the adverse effects of the estimation errors in its expected returns and covariance matrix on the performance of the portfolio. Since it is extremely challenging to accurately estimate the expected returns of an asset from its historical prices, it is a popular practice to use either a minimum variance portfolio or an optimized risk portfolio with the maximum Sharpe ratio as better proxies for the expected returns. However, due to the inherent complexity, several factors have been used to explain the expected returns.

This paper proses an algorithmic method for designing efficient portfolios by selecting stocks from the five thematic sectors of the National Stock Exchange (NSE) of India. Based on the report of the NSE on July 30, 2021, the ten most significant stocks of each of the five chosen thematic sectors are first identified [2]. Portfolios are designed for the sectors optimizing the risks and returns and exploiting their principal components. The past prices of these forty stocks for the past five years are extracted using Python from the Yahoo Finance site. To aid the portfolio construction, an LSTM model is designed for predicting future stock prices and future returns of the portfolios for different forecast horizons. Seven months after the portfolios are constructed, the actual returns and the predicted returns by the LSTM model are compared to evaluate the accuracy of the predictive model and to estimate the returns and risks associated with the thematic sectors. The thematic sectors in the NSE are based on different themes like manufacturing, services, commodities, etc. [2].

The main contribution of the current work is threefold. First, it presents two different methods of designing robust portfolios, the optimum risk portfolio, and the eigen portfolio. These portfolio design approaches are applied to five thematic sectors of stocks in the NSE. The results of the portfolios may serve as a guide to investors in the stock market for making profitable investments in the stock market. Second, a precise deep learning-based regression model is proposed exploiting the power of LSTM architecture for predicting future stock prices for robust portfolio design. Third, the returns of the portfolios highlight the current profitability of investment and the volatilities of the five thematic sectors studied in this work.

The paper is organized as follows. In Section II, some existing works on portfolio design and stock price prediction are discussed briefly. Section III presents a detailed description of the data used and the methodology followed in the work in a systematic manner. Section IV discusses the design of the LSTM model. Section V presents the results of different portfolios and the predictions of future stock prices by the LSTM models. Section VI concludes the paper.

## II. Related Work

Due to the challenging nature of the problems and their impact on real-world applications, several propositions exist in the literature for stock price prediction and robust portfolio design for optimizing returns and risk in a portfolio. The use of predictive models based on learning algorithms and deep neural net architectures for price stock price prediction is quite popular of late [3-6]. Hybrid models are also showcased that combine learning-based systems with the sentiments in the unstructured data on the social web [7-9]. The use of multi-objective optimization and eigen portfolios using principal component analysis in portfolio design has also been proposed by some researchers [10-12]. Further, genetic algorithms, fuzzy logic, and swarm intelligence are some approaches to portfolio design [13-14].

The current work presents two methods of portfolio design, the optimum risk, and the eigen portfolios, to introduce robustness while maximizing the portfolio returns for five critical thematic sectors of the NSE of India. Based on the price data from Jan 2016 to Dec 2020, ten portfolios are built, two portfolios for each sector. An LSTM model is then built for predicting the future prices of the stocks in each portfolio. Six months after the portfolio construction, the actual return for each portfolio and the return predicted by the

LSTM model are computed to analyze the profitability of each sector and the accuracy of the model.

### III. METHODOLOGY

This section highlights the seven-step research methodology that has been followed in this work. The seven steps are briefly discussed in the following.

**(1) *Choosing the sectors:*** Five thematic sectors from NIFTY of NSE, India are chosen first. The chosen sectors are as follows; (i) NIFTY services, (ii) NIFTY public sector enterprises (PSE), (iii) NIFTY multinational corporation (MNC), and (iv) NIFTY manufacturing, and (v) NIFTY commodities. Based on the criticality of stock in a particular sector, a weight is assigned to the stock which is used in deriving the aggregate sectoral index. The ten most significant stocks for each sector are chosen based on the report published by the NSE on July 30, 2021 [2].

**(2) *Data acquisition:*** For each sector, the historical prices of the ten most critical stocks are extracted using the *DataReader* function of the *data* sub-module of the *pandas_datareader* module in Python. The stock prices are extracted from the Yahoo Finance site, from Jan 1, 2016, to Dec 31, 2020. There are five features in the stock data: *open*, *high*, *low*, *close*, *volume*, and *adjusted_close*. The current work is a univariate analysis, and hence, the variable *close* is chosen as the only variable of interest.

**(3) *Derivation of the return and volatility:*** The percentage changes in the *close* values for successive days represent the daily *return* values. For computing the daily returns, the *pct_change* function of Python is used. Based on the daily returns, the daily and yearly volatilities of the ten stocks of every sector are computed. Assuming that there are 250 operational days in a calendar year, the annual volatility values for the stocks are found by multiplying the daily volatilities by a square root of 250.

**(4) *Construction of the minimum risk portfolios:*** At this step, for each sector, the minimum risk portfolio is designed. The portfolio with the minimum variance is referred to as the minimum variance portfolio. In order to identify the portfolio with the minimum variance for a given sector, first, the efficient frontier for many possible portfolios for that sector is plotted. The *efficient frontier* for a given sector represented the contour of a large number of portfolios on which the returns and the risks are plotted along the *y*-axis and the *x*-axis, respectively. The points on an efficient frontier have the property that they are the portfolios that yield the maximum return for a given risk, or they introduce the minimum risk for a given return. The left-most point on the efficient frontier depicts the minimum risk portfolio. For plotting the efficient frontier of a portfolio, weights are assigned randomly to the ten stocks over a loop which is iterated over 10,000 rounds in a Python program.

**(5) *Identifying the optimum risk portfolio:*** Minimum risk portfolios are rarely adopted in practice, and a trade-off between the risk and return is done. For optimizing the risk, Sharpe Ratio (SR) is used, as derived from (1).

$$SR = \frac{current\ portfolio\ return - riskfree\ portfolio\ return}{Current\ portfolio\ standradr\ deviation} \quad (1)$$

In other words, the Sharpe Ratio optimizes the return and the risk by yielding a substantially higher return with a very marginal increase in the risk. The portfolio with a risk of 1% is assumed to be risk-free. The portfolio with the maximum Sharpe Ratio is the optimum-risk portfolio, identified by the *idmax* function in Python.

**(6) *Eigen portfolio design:*** The principles of *principal component analysis* (PCA), an unsupervised learning approach, is used for building the eigen portfolios. PCA derives orthogonal components, which are capable of explaining the variance in the original data while reducing the dimensions. For each eigen portfolio, six components are derived for the ten stocks for a sector exploiting the PCA function of Python's *sklearn* library. For all the five sectors, the percentages of variance explained by the six components are found to be more than 80. The first component explains the highest percentage of variance, while the variances explained by the remaining components progressively decrease. Six portfolios are built for a sector based on the weights assigned to the ten stocks by the six principal components. Finally, the best eigen portfolio is identified based on the highest Sharpe Ratio it yields. The highest Sharpe Ratio portfolio is identified by using a Python function iterating over a loop and computing the weights assigned by the principal components for each eigen portfolio. Finally, the Sharpe Ratio of the portfolio is computed, and the portfolio that has the highest Sharpe Ratio is chosen as the eigen portfolio for the sector [9].

**(7) *Computation of predicted and actual portfolio returns:*** Based on the training data from January 1, 2016, to December 31, 2020, the portfolios with optimal risk are designed for all five sectors. On January 1, 2021, we create a fictitious investor who invests an amount of Indian Rupees (INR) of 100000 for each sector based on the recommendation of the optimal risk portfolio. Note that the amount of INR 100000 is just for illustrative purposes only. Our analysis will not be affected either by the currency or by the amount. To compute the future values of the stock prices and hence to predict the future value of the portfolio, we build an LSTM regression model. On Aug 2, 2021, using the LSTM model, we predict the stock prices for Aug 3, 2021. Based on the predicted stock values, we determine the predicted rate of return for each portfolio. And finally, on Aug 3, 2021, when the actual prices of the stocks are known, we determine the actual rates of return. The predicted and actual rates of return for the portfolios are compared to evaluate the profitability of the portfolios and the accuracy of the deep learning model.

### IV. THE LSTM MODEL

As explained in Section III, the stock prices are predicted with a forecast horizon of one day, using an LSTM deep learning model. This section presents the details of the architecture and the choice of various parameters in the model design. LSTM is an extended and advanced, *recurrent neural network* (RNN) having the ability in interpreting and predicting future values of sequential data like time series of stock prices or text [15]. LSTM networks maintain their state information in some specially designed memory cells or gates.

For predicting the stock prices for the next day, an LSTM model is designed and fine-tuned. The design of the model is exhibited in Fig. 1. The model uses daily *close* prices of the stock of the past 50 days as the input. The input data of 50 days with a single feature (i.e., *close* values) is represented by the data shape of (50, 1). The input layer forwards the data to the first LSTM. The LSTM layer is composed of 256 nodes. The output from the LSTM layer has a shape (50, 256). Thus, from every record in the input, the LSTM nodes extract 256 features. A dropout layer is used after the first LSTM layer that randomly switches off the output of thirty percent of the nodes to avoid model overfitting. Another LSTM layer with the same architecture as the previous one receives the output from the first and applies a dropout rate of thirty percent. A dense layer with 256 nodes receives the output from the second LSTM. The dense layer's output yields the predicted *close* price. The forecast horizon may be adjusted to different values by changing a tunable parameter. A forecast horizon of one day is used so that a prediction is made for the following day. The model is trained with a batch size of 64 and 100 epochs. With the exception of the output layer, the ReLU activation is used. The sigmoid function activation is used in the final output. The loss and the accuracy are measured using the *Huber loss function* and the *mean absolute error* function, respectively. The grid search method is used to finding out the values of the hyperparameters.

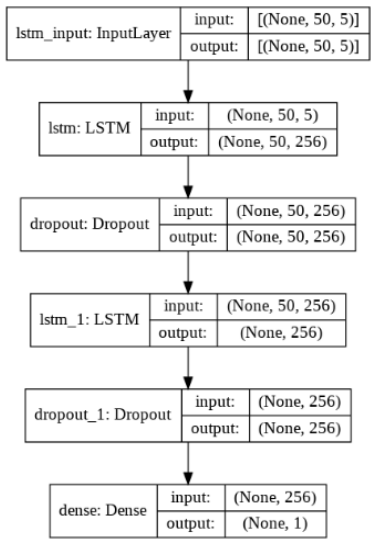

Fig. 1. The LSTM model: layers and data shapes

## V. RESULTS

This section presents the detailed results and analysis of the portfolios. The five thematic sectors of the Indian stock market we choose are (i) NIFTY services sector, (ii) NIFTY Public Sector Enterprises (PSE) sector, (iii) NIFTY Multi-National Corporations (MNC) sector, and (iv) NIFTY manufacturing sector, and (v) NIFTY commodities sector. The optimum and eigen portfolios and the LSTM predictive model are implemented in Python language with TensorFlow and Keras libraries and executed on the Google Colab platform. The model is trained and validated over 100 epochs.

***NIFTY Services sector:*** The top ten stocks and their weights for computing the index of this thematic sector based on the NSE's report on Jul 30, 2021, are the following: HDFC Bank (HDB): 14.37, Infosys (INF): 13.79, ICICI Bank (ICB): 10.91, Housing Development Finance Corp. (HDF): 10.16, Tata Consultancy Services (TCS): 7.58, Kotak Mahindra Bank (KMB): 5.61, Axis Bank (AXB): 4.21, State Bank of India (SBI): 3.83, Bajaj Finance (BJF): 3.82, and Bharti Airtel (BAL): 3.06 [2]. An imaginary investor is assumed who invested a total capital of INR 100000 on Jan 1, 2021, in the ten stocks of the NIFTY services sector. On Aug 3, 2021, the returns yielded by the two portfolio design approaches over the seven months period are computed. The LSTM model is also used for predicting the stock prices and the return one day in advance. Hence, the LSTM predicted return is computed on Aug 2, 2021, and then on Aug 3, the predicted return is compared with the optimum risk portfolio's return. Tables I – III depict the returns of the optimum risk and the eigen portfolios, and the LSTM model predicted return for an investor who invested following the allocations suggested by the optimum risk portfolio. Fig. 2 depicts the efficient frontier, the portfolios with minimum risk, and optimum risk for the NIFTY services sector. As an illustration, Fig 3 depicts the plot of actual prices vs. corresponding predicted prices of the most significant stock in this sector, HDFC Bank, from Jan 1, 2021, to Aug 2, 2021.

TABLE I. OPT RISK PORTFOLIO RETURN (NIFTY SERVICES)

| Stock | Weights | Date: Jan 1, 2021 | | | Date: Aug 3, 2021 | | ROI |
|---|---|---|---|---|---|---|---|
| | | Amount Invested | Actual Price | No of Stock | Actual Price | Actual Value | |
| HDB | 0.15119 | 15119 | 1425 | 10.61 | 1435 | 15225 | |
| INF | 0.17252 | 17252 | 1260 | 13.69 | 1655 | 22660 | |
| ICB | 0.02400 | 2400 | 528 | 4.55 | 690 | 3136 | |
| HDF | 0.00738 | 738 | 2569 | 0.29 | 2555 | 734 | |
| TCS | 0.16375 | 16375 | 2928 | 5.59 | 3285 | 18372 | 15.56 |
| KMB | 0.08380 | 8380 | 1994 | 4.20 | 1686 | 7086 | |
| AXB | 0.01450 | 1450 | 624 | 2.32 | 739 | 1717 | |
| SBI | 0.04352 | 4352 | 279 | 15.60 | 447 | 6973 | |
| BJF | 0.19777 | 19777 | 5280 | 3.75 | 6332 | 23717 | |
| BAL | 0.14157 | 14157 | 515 | 27.49 | 580 | 15944 | |
| Total | | 100000 | | | | 115564 | |

TABLE II. EIGEN PORTFOLIO RETURN (NIFTY SERVICES)

| Stock | Weights | Date: Jan 1, 2021 | | | Date: Aug 3, 2021 | | ROI |
|---|---|---|---|---|---|---|---|
| | | Amount Invested | Actual Price | No of Stocks | Actual Price | Actual Value | |
| HDB | 0.1057 | 10570 | 1425 | 7.42 | 1435 | 10644 | |
| INF | 0.0577 | 5770 | 1260 | 4.58 | 1655 | 7579 | |
| ICB | 0.1323 | 13230 | 528 | 25.05 | 690 | 17289 | |
| HDF | 0.1199 | 11990 | 2569 | 4.67 | 2555 | 11925 | |
| TCS | 0.0458 | 4580 | 2928 | 1.56 | 3285 | 5138 | 18.25 |
| KMB | 0.1021 | 10210 | 1994 | 5.12 | 1686 | 8633 | |
| AXB | 0.1162 | 11620 | 624 | 18.62 | 739 | 13762 | |
| SBI | 0.1348 | 13480 | 279 | 48.32 | 447 | 21597 | |
| BJF | 0.1087 | 10870 | 5280 | 2.06 | 6332 | 13036 | |
| BAL | 0.0767 | 7680 | 515 | 14.91 | 580 | 8649 | |
| Total | | 100000 | | | | 118252 | |

TABLE III. LSTM PREDICTED RETURN (NIFTY SERVICES)

| Stock | Date: Aug 3, 2021 | | | ROI |
|---|---|---|---|---|
| | Predicted Price | No of Stocks | Predicted Value | |
| HDB | 1436 | 10.61 | 15236 | |
| INF | 1631 | 13.69 | 22328 | |
| ICB | 689 | 4.55 | 3135 | |
| HDF | 2448 | 0.29 | 710 | |
| TCS | 3250 | 5.59 | 18168 | 14.14 |
| KMB | 1683 | 4.20 | 7069 | |
| AXB | 722 | 2.32 | 1675 | |
| SBI | 422 | 15.60 | 6583 | |
| BJF | 6343 | 3.75 | 23786 | |
| BAL | 562 | 27.49 | 15449 | |
| Total | | | 114139 | |

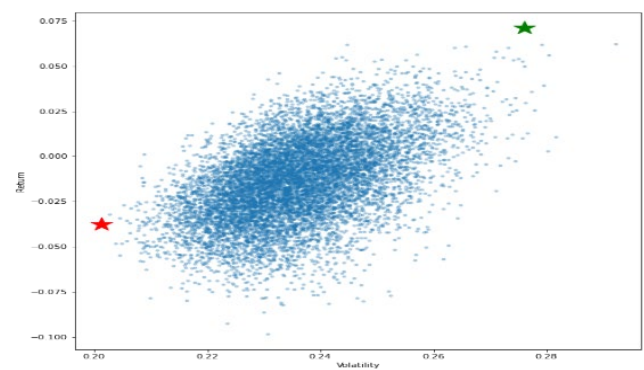

Fig. 2. The efficient frontier of the NIFTY services sector was built on Jan 1, 2021. The red star and green star depict the min. risk and the opt. risk portfolios, respectively.

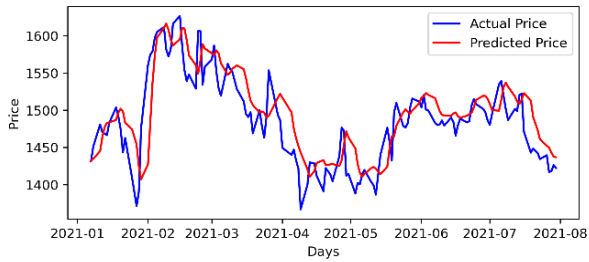

Fig. 3. Actual vs. predicted values of the HDFC Bank (HDB) stock as predicted by the LSTM model (Period: Jan 1, 2021, to Aug 2, 2021)

TABLE IV.  OPT RISK PORTFOLIO RETURN (NIFTY PSE)

| Stock | Weights | Date: Jan 1, 2021 | | | Date: Aug 3, 2021 | | ROI |
| --- | --- | --- | --- | --- | --- | --- | --- |
| | | Amount Invested | Actual Price | No of Stock | Actual Price | Actual Value | |
| PGC | 0.07837 | 7837 | 142 | 55.19 | 174 | 9603 | |
| NTP | 0.02478 | 2478 | 99 | 25.03 | 118 | 2954 | |
| ONG | 0.05700 | 5700 | 93 | 61.29 | 118 | 7232 | |
| BPC | 0.15816 | 15816 | 382 | 41.40 | 463 | 19170 | |
| CIL | 0.00299 | 299 | 135 | 2.21 | 144 | 319 | 51.68 |
| IOC | 0.00085 | 85 | 92 | 0.92 | 106 | 98 | |
| GAI | 0.06034 | 6034 | 124 | 48.66 | 143 | 6959 | |
| BEL | 0.11891 | 11891 | 126 | 94.37 | 182 | 17176 | |
| NMD | 0.14470 | 14470 | 116 | 124.74 | 181 | 22578 | |
| SAI | 0.35390 | 35390 | 75 | 471.87 | 139 | 65590 | |
| Total | | 100000 | | | | 151679 | |

TABLE V.  EIGEN PORTFOLIO RETURN (NIFTY PSE)

| Stock | Weights | Date: Jan 1, 2021 | | | Date: Aug 3, 2021 | | ROI |
| --- | --- | --- | --- | --- | --- | --- | --- |
| | | Amount Invested | Actual Price | No of Stock | Actual Price | Actual Value | |
| PGC | 0.0786 | 7860 | 142 | 55.35 | 174 | 9631 | |
| NTP | 0.0895 | 8950 | 99 | 90.40 | 118 | 10668 | |
| ONG | 0.1046 | 10460 | 93 | 112.47 | 118 | 13272 | |
| BPC | 0.1130 | 11300 | 382 | 29.58 | 463 | 13696 | |
| CIL | 0.0842 | 8420 | 135 | 62.37 | 144 | 8981 | 32.67 |
| IOC | 0.1205 | 12050 | 92 | 130.98 | 106 | 13884 | |
| GAI | 0.0903 | 9030 | 124 | 72.82 | 143 | 10414 | |
| BEL | 0.0980 | 9800 | 126 | 77.78 | 182 | 14156 | |
| NMD | 0.1041 | 10410 | 116 | 89.74 | 181 | 16243 | |
| SAI | 0.1172 | 11720 | 75 | 156.27 | 139 | 21721 | |
| Total | | 100000 | | | | 132666 | |

TABLE VI.  LSTM PREDICTED RETURN (NIFTY PSE)

| Stock | Date: Aug 3, 2021 | | | ROI |
| --- | --- | --- | --- | --- |
| | Predicted Price | No of Stocks | Predicted Value | |
| PGC | 169 | 55.19 | 9327 | |
| NTP | 117 | 25.03 | 2929 | |
| ONG | 117 | 61.29 | 7171 | |
| BPC | 454 | 41.40 | 18796 | |
| CIL | 143 | 2.21 | 316 | 47.78 |
| IOC | 103 | 0.92 | 95 | |
| GAI | 142 | 48.66 | 6910 | |
| BEL | 182 | 94.37 | 17175 | |
| NMD | 175 | 124.74 | 21830 | |
| SAI | 134 | 471.87 | 63231 | |
| Total | | | 147780 | |

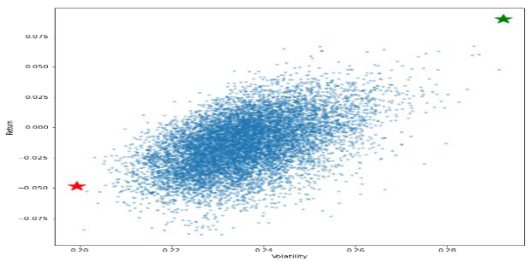

Fig. 4. The efficient frontier of the NIFTY PSE sector was built on Jan 1, 2021. The red star and the green star depict the min. risk and the opt. risk portfolios, respectively.

*NIFTY PSE sector:* The top ten stocks of the NIFTY PSE and their weights are: Power Grid Corp. of India (PGC): 12.82, NTPC (NTP): 12.31, Oil & Natural Gas Corp. (ONG): 9.22, Bharat Petroleum Corp. (BPC): 9.11, Coal India (CIL): 6.58, Indian Oil Corp. (IOC): 5.75, GAIL (GAI): 5.43, Bharat Electronics (BEL): 4.83, NMDC (NMD): 4.54, and Steel Authority of India (SAI): 4.50 [2]. Tables IV-VI show the returns of the optimum risk and the eigen portfolios, and the LSTM model for a period of seven months, Jan 1, 2021, to Aug 1, 2021. Fig. 4 shows the efficient frontier, while Fig. 5 displays the actual-vs-predicted prices of Power Grid Corporation (PGC), which is the leading stock of the oil & gas sector.

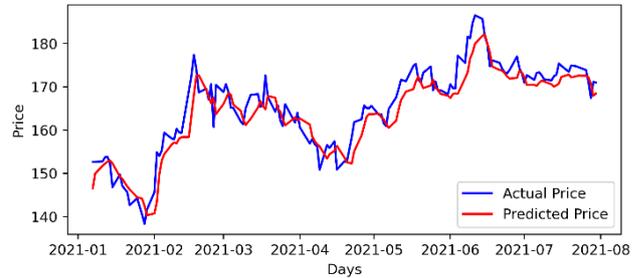

Fig. 5. Actual vs. predicted values of Power Grid Corporation (PGC) stock as predicted by the LSTM model (Period: Jan 1, 2021, to Aug 2, 2021)

TABLE VII.  OPT RISK PORTFOLIO RETURN (NIFTY MNC)

| Stock | Weights | Date: Jan 1, 2021 | | | Date: Aug 3, 2021 | | ROI |
| --- | --- | --- | --- | --- | --- | --- | --- |
| | | Amount Invested | Actual Price | No of Stock | Actual Price | Actual Value | |
| NIL | 0.21102 | 21102 | 18451 | 1.14 | 18284 | 20911 | |
| HUL | 0.28057 | 28057 | 2388 | 11.75 | 2387 | 28045 | |
| MSL | 0.08052 | 8052 | 7691 | 1.05 | 7199 | 7537 | |
| BIL | 0.19814 | 19814 | 3568 | 5.55 | 3573 | 19842 | |
| VDL | 0.00169 | 169 | 160 | 1.06 | 313 | 331 | 9.85 |
| ACL | 0.05224 | 5224 | 251 | 20.81 | 422 | 8783 | |
| CPL | 0.04145 | 4145 | 1578 | 2.63 | 1674 | 4397 | |
| MPL | 0.07173 | 7173 | 1530 | 4.69 | 2692 | 12621 | |
| ALL | 0.01187 | 1187 | 99 | 11.99 | 141 | 1691 | |
| USL | 0.05077 | 5077 | 582 | 8.72 | 653 | 5696 | |
| Total | 1.00000 | 100000 | | | | 109854 | |

TABLE VIII  EIGEN PORTFOLIO RETURN (NIFTY MNC)

| Stock | Weights | Date: Jan 1, 2021 | | | Date: Aug 3, 2021 | | ROI |
| --- | --- | --- | --- | --- | --- | --- | --- |
| | | Amount Invested | Actual Price | No of Stock | Actual Price | Actual Value | |
| NIL | 0.0916 | 9160 | 18451 | 0.50 | 18284 | 9077 | |
| HUL | 0.1072 | 10720 | 2388 | 4.49 | 2387 | 10716 | |
| MSL | 0.1220 | 12200 | 7691 | 1.59 | 7199 | 11420 | |
| BIL | 0.1210 | 12100 | 3568 | 3.39 | 3573 | 12117 | |
| VDL | 0.1149 | 11490 | 160 | 71.81 | 313 | 22477 | 26.67 |
| ACL | 0.1240 | 12400 | 251 | 49.40 | 422 | 20848 | |
| CPL | 0.0918 | 9180 | 1578 | 5.82 | 1674 | 9738 | |
| MPL | 0.0271 | 2710 | 1530 | 1.77 | 2692 | 4768 | |
| ALL | 0.1000 | 10000 | 99 | 101.01 | 141 | 14242 | |
| USL | 0.1004 | 10040 | 582 | 17.25 | 653 | 11265 | |
| Total | 1.0000 | 100000 | | | | 126668 | |

TABLE IX.  LSTM PREDICTED RETURN (NIFTY MNC)

| Stock | Date: Aug 3, 2021 | | | ROI |
| --- | --- | --- | --- | --- |
| | Predicted Price | No of Stocks | Predicted Value | |
| NIL | 18287 | 1.14 | 20847 | |
| HUL | 2364 | 11.75 | 27777 | |
| MSL | 7110 | 1.05 | 7466 | |
| BIL | 3458 | 5.55 | 19192 | |
| VDL | 298 | 1.06 | 316 | 7.92 |
| ACL | 401 | 20.81 | 8345 | |
| CPL | 1708 | 2.63 | 4492 | |
| MPL | 2615 | 4.69 | 12264 | |
| ALL | 131 | 11.99 | 1571 | |
| USL | 648 | 8.72 | 5651 | |
| Total | | | 107921 | |

*NIFTY MNC sector:* The ten stocks that have the most significant contributions on the computation of the index of this thematic sector are as follows: Nestle India (NIL): 9,97, Hindustan Unilever (HUL): 9.39, Maruti Suzuki India (MSL): 8.99, Britannia Industries (BIL): 7.74, Vedanta (VDL): 7.31, Ambuja Cements (ACL): 5.79, Colgate Palmolive India (COL): 4.35, MphasiS (MPL): 4.10, Ashok Leyland (ALL): 3.66, and United Spirits (USL): 3.66 [2]. Tables VII-IX depict the returns from the optimum risk and eigen portfolios, and the LSTM model for the period Jan 1, 2021, to Aug 2, 2021. Fig. 6 and Fig 7 depict the efficient frontier and the plot of actual vs. predicted prices of the leading stock of the sector, Nestle India (NIL).

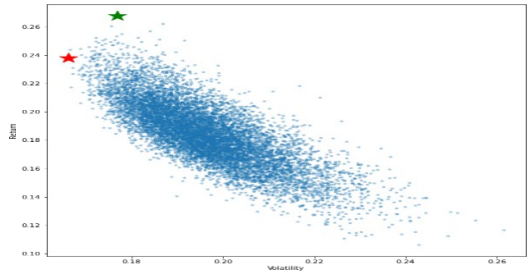

Fig. 6. The efficient frontier of the NIFTY MNC sector was built on Jan 1, 2021. The red star and the green star depict the min. risk and the opt. risk portfolios, respectively.

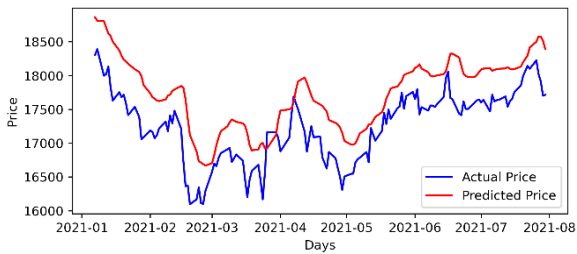

Fig. 7. Actual vs. predicted values of the Nestle India (NIL) stock as predicted by the LSTM model (Period: Jan 1, 2021, to Aug 2, 2021)

TABLE X. OPT RISK PORTFOLIO RETURN (NIFTY MFG)

| Stock | Weights | Date: Jan 1, 2021 | | | Date: Aug 3, 2021 | | ROI |
|---|---|---|---|---|---|---|---|
| | | Amount Invested | Actual Price | No of Stock | Actual Price | Actual Value | |
| TSL | 0.03060 | 3060 | 643 | 4.76 | 1407 | 6696 | |
| SPI | 0.11212 | 11212 | 596 | 18.81 | 795 | 14956 | |
| RIL | 0.24655 | 24655 | 1988 | 12.40 | 2088 | 25895 | |
| JSL | 0.13097 | 13097 | 390 | 33.58 | 740 | 24851 | |
| MSL | 0.04237 | 4237 | 7691 | 0.55 | 7199 | 3966 | 28.09 |
| HIL | 0.00814 | 814 | 238 | 3.42 | 449 | 1536 | |
| DVL | 0.25772 | 25772 | 3849 | 6.70 | 5007 | 33526 | |
| DRL | 0.13878 | 13878 | 5241 | 2.65 | 4722 | 12504 | |
| MML | 0.02490 | 2490 | 732 | 3.40 | 771 | 2623 | |
| VDL | 0.00785 | 785 | 160 | 4.91 | 313 | 1536 | |
| Total | | 100000 | | | | 128089 | |

TABLE XI. EIGEN PORTFOLIO RETURN (NIFTY MFG)

| Stock | Weights | Date: Jan 1, 2021 | | | Date: Aug 3, 2021 | | ROI |
|---|---|---|---|---|---|---|---|
| | | Amount Invested | Actual Price | No of Stock | Actual Price | Actual Value | |
| TSL | 0.1366 | 13660 | 643 | 21.24 | 1407 | 29891 | |
| SPI | 0.0753 | 7530 | 596 | 12.63 | 795 | 10044 | |
| RIL | 0.0803 | 8030 | 1988 | 4.04 | 2088 | 8434 | |
| JSL | 0.1212 | 12120 | 390 | 31.08 | 740 | 22997 | |
| MSL | 0.0985 | 9850 | 7691 | 1.28 | 7199 | 9220 | 55.85 |
| HIL | 0.1294 | 12940 | 238 | 54.37 | 449 | 24412 | |
| DVL | 0.0710 | 7100 | 3849 | 1.84 | 5007 | 9236 | |
| DRL | 0.0626 | 6260 | 5241 | 1.19 | 4722 | 5640 | |
| MML | 0.0892 | 8920 | 732 | 12.19 | 771 | 9395 | |
| VDL | 0.1359 | 13590 | 160 | 84.94 | 313 | 26585 | |
| Total | | 100000 | | | | 155854 | |

TABLE XII. LSTM PREDICTED RETURN (NIFTY MFG)

| Stock | Date: Aug 3, 2021 | | | ROI |
|---|---|---|---|---|
| | Predicted Price | No of Stocks | Predicted Value | |
| TSL | 1387 | 4.76 | 6602 | |
| SPI | 740 | 18.81 | 13919 | |
| RIL | 2053 | 12.40 | 25457 | |
| JSL | 722 | 33.58 | 24245 | |
| MSL | 7110 | 0.55 | 3911 | 25.68 |
| HIL | 428 | 3.42 | 1464 | |
| DVL | 5009 | 6.70 | 33560 | |
| DRL | 4733 | 2.65 | 12543 | |
| MML | 741 | 3.40 | 2519 | |
| VDL | 298 | 4.91 | 1463 | |
| Total | | | 125683 | |

*NIFTY Manufacturing sector:* The top ten stocks of NIFTY manufacturing are: Tata Steel (TSL): 5.62, Sun Pharmaceutical Industries (SPI): 4.34, Reliance Industries (RIL): 4.25, JSW Steel (JSL): 3.70, Maruti Suzuki India (MSL): 3.52, Hindalco Industries (HIL): 3,37, Divi's Labs (DVL): 3.24, Dr. Reddy's Labs (RDL): 2.97, Mahindra and Mahindra (MML): 2.70, and Vedanta (VDL): 2.62 [2]. Tables X-XII exhibit the results of the portfolios and the LSTM predicted returns of this sector. Fig. 8 and Fig 9 depict the efficient frontier and the plot of actual vs. predicted prices of the leading stock of the sector, Tata Steel (TSL).

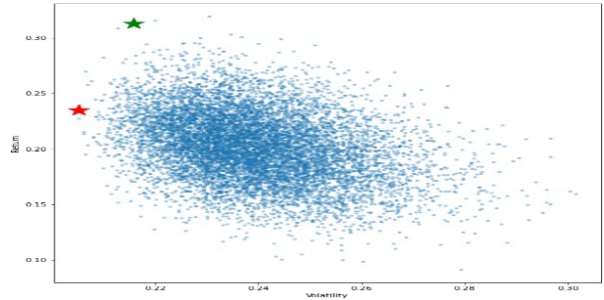

Fig. 8. The efficient frontier of the NIFTY manufacturing sector was built on Jan 1, 2021. The red star and the green star depict the min. risk and the opt. risk portfolios, respectively.

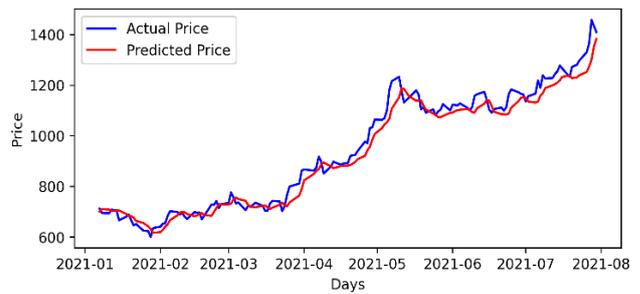

Fig. 9. Actual vs. predicted values of the Tata Steel (TSL) stock as predicted by the LSTM model (Period: Jan 1, 2021, to Aug 2, 2021)

*NIFTY Commodities sector:* The top ten stocks of NIFTY commodities are: Tata Steel (TSL): 9.89, Reliance Industries (RIL): 9.50, UltraTech Cement (UTC): 7.76, JSW Steel (JSL): 6.29, Hindalco Industries (HIL): 5.73, Grasim Industries (GRI): 5.22, NTPC (NTP): 4.95, UPL (UPL): 3.92, Oil & Natural Gas Corp. (ONG): 3.71, and Bharat Petroleum Corp. (BPC): 3.67 [2]. Tables XIII-XV exhibit the results of the portfolios and the LSTM predicted returns of this sector. Fig. 10 depicts the actual-vs-predicted prices plot of the second most significant stock of the sector, Reliance Industries (RIL). Note that the plot of the most significant stock of this sector, Tata Steel is depicted in Fig 9 while discussing the NIFTY manufacturing sector.

TABLE XIII. OPT RISK PORTFOLIO RETURN (NIFTY COMD)

| Stock | Weights | Date: Jan 1, 2021 | | | Date: Aug 3, 2021 | | ROI |
|---|---|---|---|---|---|---|---|
| | | Amount Invested | Actual Price | No of Stock | Actual Price | Actual Value | |
| TSL | 0.00105 | 105 | 643 | 0.16 | 1410 | 230 | |
| RIL | 0.28610 | 28610 | 1988 | 14.39 | 2073 | 29833 | |
| UTC | 0.10166 | 10166 | 5291 | 1.92 | 7642 | 14683 | |
| JSL | 0.23774 | 23774 | 390 | 60.96 | 747 | 45536 | |
| HIL | 0.11765 | 11765 | 238 | 49.43 | 446 | 22047 | 46.49 |
| GRI | 0.04617 | 4617 | 933 | 4.95 | 1592 | 7878 | |
| NTP | 0.11367 | 11367 | 99 | 114.82 | 118 | 13549 | |
| UPL | 0.03152 | 3152 | 469 | 6.72 | 791 | 5316 | |
| ONG | 0.00717 | 717 | 93 | 7.71 | 118 | 910 | |
| BPC | 0.05727 | 5727 | 382 | 14.99 | 458 | 6866 | |
| Total | | 100000 | | | | 146488 | |

TABLE XIV. EIGEN PORTFOLIO RETURN (NIFTY COMD)

| Stock | Weights | Date: Jan 1, 2021 | | | Date: Aug 3, 2021 | | ROI |
| --- | --- | --- | --- | --- | --- | --- | --- |
| | | Amount Invested | Actual Price | No of Stock | Actual Price | Actual Value | |
| TSL | 0.1285 | 12850 | 643 | 19.98 | 1410 | 28178 | |
| RIL | 0.0786 | 7860 | 1988 | 3.95 | 2073 | 8196 | |
| UTC | 0.1141 | 11410 | 5291 | 2.16 | 7642 | 16480 | |
| JSL | 0.1157 | 11570 | 390 | 29.67 | 747 | 22161 | |
| HIL | 0.1173 | 11730 | 238 | 49.29 | 446 | 21981 | 60.70 |
| GRI | 0.1063 | 10630 | 933 | 11.39 | 1592 | 18138 | |
| NTP | 0.0777 | 7770 | 99 | 78.48 | 118 | 9261 | |
| UPL | 0.0880 | 8800 | 469 | 18.76 | 791 | 14842 | |
| ONG | 0.0889 | 8890 | 93 | 95.59 | 118 | 11280 | |
| BPC | 0.0849 | 8490 | 382 | 22.23 | 458 | 10179 | |
| Total | | 100000 | | | | 160696 | |

TABLE XV. LSTM PREDICTED RETURN (NIFTY COMD)

| Stock | Date: Aug 3, 2021 | | | ROI |
| --- | --- | --- | --- | --- |
| | Predicted Price | No of Stocks | Predicted Value | |
| TSL | 1387 | 0.16 | 222 | |
| RIL | 2053 | 14.39 | 29543 | |
| UTC | 7557 | 1.92 | 14509 | |
| JSL | 722 | 60.96 | 44013 | |
| HIL | 428 | 49.43 | 21156 | 43.46 |
| GRI | 1524 | 4.95 | 7544 | |
| NTP | 117 | 114.82 | 13434 | |
| UPL | 794 | 6.72 | 5336 | |
| ONG | 117 | 7.71 | 902 | |
| BPC | 454 | 14.99 | 6805 | |
| Total | | | 143464 | |

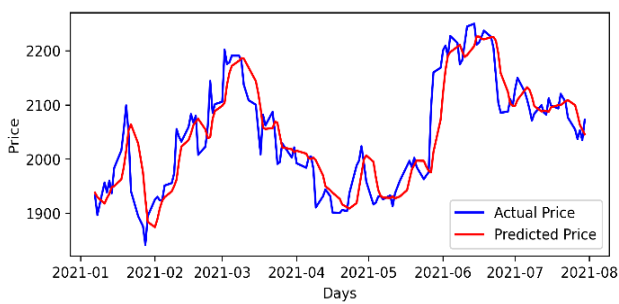

Fig. 10. Actual vs. predicted values of the Reliance Industries (RIL) stock as predicted by the LSTM model (Period: Jan 1, 2021, to Aug 2, 2021)

*Summary results:* Table XVI presents a summary of the results of the returns of the ten portfolios and the returns based on the predictions by the LSTM model.

For the NIFTY services sector, the eigen portfolio's return is marginally higher than that of the optimum portfolio. The stocks which gained at prices for this sector are SBI (60.21%), INF (31.35%), and ICB (30.68%), while the top losers are KMB (-15.44%) and HDF (-0.54%). While the eigen portfolio assigned higher weights to SBI and ICB, the stock of INF received a higher weight from the optimum portfolio. For the top losing stocks, KMB and HDF received higher weights from the eigen portfolio. Due to these allocations, in some cases, the eigen portfolio has gained over the optimum portfolio while losing in some others. The result is a narrow difference in the overall returns of the two portfolios. However, the return of the eigen portfolio is marginally higher. While all the stocks under the NIFTY PSE sector had their prices increased during the period Jan 1 – Aug 2, 2021, the top three gainers were SAI (85.33%), NMD (56.03%), and BEL (44.44%). The optimum portfolio allocated higher weights to all these three stocks in comparison to the eigen portfolio, resulting in a substantially higher return for the NIFTY PSE sector for the optimal portfolio. The returns of the eigen portfolio are higher than those of the optimum portfolio for the MNC and manufacturing sectors because it allocated higher weights to the top gainer VDL(95.63%). Finally, for the commodities sector, the higher return from the eigen portfolio is due to its increased allocation of weight to the top gainer TSL (119.28%). Among the five sectors, except for the services sector, eigen portfolios yielded a higher return than their corresponding optimum risk portfolios.

TABLE XVI. THE SUMMARY OF THE RESULTS

| Portfolio | Opt. Portfolio Return (%) | Eigen Portfolio Return (%) | LSTM Predicted Return (%) |
| --- | --- | --- | --- |
| NIFTY Services | 15.56 | 18.25 | 14.14 |
| NIFTY PSE | 51.68 | 32.67 | 47.78 |
| NIFTY MNC | 9.85 | 26.67 | 7.92 |
| NIFTY Manufacturing | 28.09 | 55.85 | 25.68 |
| NIFTY Commodities | 46.49 | 60.70 | 43.46 |

## VI. CONCLUSION

We have presented optimized risk portfolios and eigen portfolios for five thematic sectors of NSE, India, taking into account the ten most significant stocks from those sectors using their historical prices from Jan 1, 2016, to Dec 31, 2020. An LSTM model is also designed for predicting stock prices with a forecast horizon of one day. After a hold-out period of seven months, the actual and the predicted return of the portfolios are computed. The returns yielded by the eigen portfolios are found to be higher than those of the optimum risk portfolios for all the sectors except for the NIFTY services sector. The accuracy of the LSTM model is found to be very high as for all the sectors, the predicted return by the model was found to be very close to the actual returns from the optimum risk portfolios.

## REFERENCES

[1] H. Markowitz, "Portfolio selection", *Journ. of Fin.*, vol 7, no. 1, pp. 77-91, 1952.

[2] NSE Website: http://www1.nseindia.com

[3] J. Sen and S. Mehtab, "Accurate stock price forecasting using robust and optimized deep learning model", *Proc. of IEEE CONIT*, pp.1-9, . 2021.

[4] S. Mehtab and J. Sen, "Stock price prediction using convolutional neural network on a multivariate time series", *Proc. of 3rd NCMLAI' 20*, Feb 2020.

[5] W. Jiang, "Applications of deep learning in stock market prediction: Recent progress", *Expert Sys with Appl*, vol, 184, art. Id: 115537, Dec 2021.

[6] J. Sen, "Stock price prediction using machine learning and deep learning frameworks", *Proc. of ICBAI*, Dec 20-22, 2018, India.

[7] M. Li, C. Yang, J. Zhang, D. Puthal, Y. Luo, and J. Li, "Stock market analysis using social networks", *Proc. of ACSW'18*, pp. 1-10, Jan 2018.

[8] S. Metab and J. Sen, "A robust predictive model for stock price prediction using deep learning and natural language processing", *Proc. of 7th BAICONF'190*, Dec 5 -7, 2019.

[9] C-H. Chao, _-H. Ting, T-H. Tsai, and M-C. Chen, "Opinion mining and the visualization of stock selection in quantitative trading", *Proc. of TAAI*, pp. 1-6, Kaohsiung, Taiwan, Jan 16, 2020.

[10] J. Sen and S. Mehtab, "Optimum risk portfolio and eigen portfolio: a comparative analysis using selected stocks from the Indian stock market, *Int. J. of Bus. Forcstng and Mktg. Intel.*, Inderscience Publishers, 2021 (In press).

[11] C. Chen and Y. Zhou, "Robust multi-objective portfolio with higher moments", *Expert Sys with Appl.*, vol. 100, pp. 165-181, 2018.

[12] Y. Peng, P. H. M. Albuquerque, I-F, do Nascimento, and J. V. F. Machado, "Between nonlinearities, complexity, and noises: an application on portfolio selection using kernel principal component analysis", *Entropy*, vol 21, no 4, p. 376, 2019.

[13] K. Erwin and A. Engelbrecht, "Improved set-based particle swarm optimization for portfolio optimization," *Proc. of IEEE SSCI*, pp.1573-1580, Canberra, Australia, Dec 1-4, 2020.

[14] Z. Wang, X. Zhang, Z. Zhang, and D. Sheng, "Credit portfolio optimization: a multi-objective genetic algorithm approach", *Bora Istanbul Review*, Jan 2021. (In press).

[15] O. Campesato, *Angular and Deep Learning*, Mercury Learning & Information, 2020, ISBN-13: 978-1683924739, Herndon, VA, USA.